Research article

Paola Prete*, Daniel Wolf, Fabio Marzo and Nico Lovergine

# Nanoscale spectroscopic imaging of GaAs-AlGaAs quantum well tube nanowires: correlating luminescence with nanowire size and inner multishell structure

**Abstract:** The luminescence and inner structure of GaAs-AlGaAs quantum well tube (QWT) nanowires were studied using low-temperature cathodoluminescence (CL) spectroscopic imaging, in combination with scanning transmission electron microscopy (STEM) tomography, allowing for the first time a robust correlation between the luminescence properties of these nanowires and their size and inner 3D structure down to the nanoscale. Besides the core luminescence and minor defects-related contributions, each nanowire showed one or more QWT peaks associated with nanowire regions of different diameters. The values of the GaAs shell thickness corresponding to each QWT peak were then determined from the nanowire diameters by employing a multishell growth model upon validation against experimental data (core diameter and GaAs and AlGaAs shell thickness) obtained from the analysis of the 3D reconstructed STEM tomogram of a GaAs-AlGaAs QWT nanowire. We found that QWT peak energies as a function of thus-estimated (3–7 nm) GaAs shell thickness are 40–120 meV below the theoretical values of exciton recombination for uniform QWTs symmetrically wrapped around a central core. However, the analysis of the 3D tomogram further evidenced azimuthal asymmetries as well as (azimuthal and axial) random fluctuations of the GaAs shell thickness, suggesting that the red-shift of QWT emissions is prominently due to carrier localization. The CL mapping of QWT emission intensities along the nanowire axis allowed to directly image the nanoscale localization of the emission, supporting the above picture. Our findings contribute to a deeper understanding of the luminescence-structure relationship in QWT nanowires and will foster their applications as efficient nanolaser sources for future monolithic integration onto silicon.

## 1 Introduction

Nanowires of III–V compound semiconductors have attracted considerable research interest in recent years due to their potential applications in novel nanophotonic devices, such as nanolasers [1], photodetectors [2, 3], and efficient nanowire array solar cells [4]. The radial modulation of the nanowire composition to form core-shell heterostructures impacts the design of such nanodevices by adding new degrees of freedom associated with quantum confinement. In the case of GaAs-AlGaAs core-multishell nanowires, the insertion of a few-nanometer-thin GaAs shell in between two relatively thicker AlGaAs shells overgrown around the central GaAs core may form quantum-confined electron and hole states inside the thin-wrapped (around the core) GaAs shell, leading to what is called a quantum well nanowire [5] or a quantum well tube (QWT) [6, 7] structure. In particular, single and multiple QWT structures within GaAs-AlGaAs core-multishell nanowires have been found to show optically pumped lasing [5, 8, 9], with improved gain and low threshold with respect to bulk nanowire counterparts [10]. Therefore, they represent potential candidates as efficient nanolaser sources for direct monolithic integration onto silicon.

In this respect, a detailed knowledge of the core-multishell nanowire optical (radiative) properties and correlation of the observed emission channels with their inner nanoscale structures is a necessary step to fully

*Corresponding author: Paola Prete,* Istituto per la Microelettronica e Microsistemi del CNR, SS Lecce, Via Monteroni, I-73100 Lecce, Italy, e-mail: paola.prete@cnr.it.
https://orcid.org/0000-0002-4948-4718
**Daniel Wolf:** Institute for Solid State Research, Leibniz Institute for Solid State and Materials Research, Helmholtzstrasse 20, D-01069 Dresden, Germany. https://orcid.org/0000-0001-5000-8578
**Fabio Marzo and Nico Lovergine:** Dipartimento di Ingegneria dell'Innovazione, Università del Salento, Via Monteroni, I-73100 Lecce, Italy. https://orcid.org/0000-0003-0190-4899 (N. Lovergine) understand the origin of nanowire luminescence emissions and the nature of confined carrier states within these complex nanostructures. To date, studies in the literature [7, 11, 12] have employed a combination of micro-photoluminescence (μ-PL) measurements, performed on single nanowires, and transmission electron microscopy (TEM) observations of suitably prepared nanowire cross-sections to ascribe observed luminescence emissions to the nanowire inner structure (i.e. QWT morphology, shell thicknesses, and GaAs/AlGaAs interface roughness). However, the limited spatial resolution of μ-PL, along with the inevitable nanowire-to-nanowire variability within a given sample, may prevent the precise correlation between the nanowire emission channels and its actual size and inner multishell structure.

In this work, we report for the first time the direct correlation between the nanoscale luminescence properties of GaAs-AlGaAs QWT nanowires and their size variation and related inner 3D structure. The study was carried out using low-temperature cathodoluminescence (CL) spectroscopy and imaging performed on single nanowires in a high-resolution scanning electron microscope in combination with scanning TEM (STEM) tomography. The latter was used to reconstruct the detailed 3D morphology and GaAs shell thickness distribution of a typical QWT nanowire. These findings were employed to validate a novel multishell growth model able to predict (within a few-percent precision) the GaAs and AlGaAs shell thicknesses within the observed nanowires, obtaining a robust correlation between energies of QWT emissions and their actual thickness.

## 2 Experimental

Nominally undoped GaAs-AlGaAs core-multishell nanowires were grown on semi-insulating (111)B-GaAs substrates using an Aixtron RD200 MOVPE reactor operated at 50 mbar pressure under ultra-pure $H_2$ carrier gas. Trimethylgallium, trimethylaluminum ($Me_3Al$), and tertiarybutylarsine ($tBuAsH_2$) were used as Ga, Al, and As precursors, respectively. Vertically aligned GaAs nanowires were grown at 400°C by the Au-catalyzed (so-called vapor-liquid-solid or VLS [13]) method in the form of dense arrays using colloidal Au nanoparticles following the procedures reported in Ref. [14], obtaining areal density on the substrate in the $10^6$–$10^8$ cm$^{-2}$ range. GaAs nanowires were radially overgrown at 650°C by a first AlGaAs shell (for 2 min) followed by a GaAs shell (Sample A for 1 min and Sample B for 2 min), a second AlGaAs shell (for 3 min), and a final GaAs cap layer (for 1 min), with the latter to avoid the oxidation of the AlGaAs alloy when exposed to ambient atmosphere. Within such nanostructures, the central GaAs shell acts as a potential well, whereas the two AlGaAs shells act as confining barriers for both electrons and holes. Both the core and shell materials were grown using a precursor V/III ratio in the vapor of 20:1, whereas the $Me_3Al$/group III precursor ratio in the vapor during AlGaAs shell growth was kept at $x=0.50$; low-temperature PL and energy-dispersive X-ray spectroscopy performed in a TEM have shown that these conditions allowed to incorporate an Al molar fraction xAl = 0.33 into the AlGaAs shells [15, 16]. To avoid the broadening of GaAs/AlGaAs heterointerfaces due to vapor intermixing, the supply of group III species was turned off after each shell growth and the reactor was flushed with $tBuAsH_2 + H_2$ for 1 min before resuming the growth.

The morphology and size (length and diameters) of as-grown core-multishell nanowires were assessed by field-emission scanning electron microscopy (FE-SEM) using a Zeiss Sigma VP microscope equipped with a high-resolution FE-SEM Gemini electron column operated with a primary electron beam energy of 5 keV and using the Gemini "in-lens" secondary electron detector for best spatial resolution.

The spectroscopic characterization and nanoscale imaging of the nanowire radiative emission were performed at 7 K by means of CL measurements inside the Zeiss Sigma VP microscope using a Gatan MonoCL4 system and a liquid-helium cooling stage assembly Gatan CF302. In this case, the Gemini column was operated in high current mode and with a primary beam energy of 5–10 keV. CL measurements were performed on single nanowires: to this purpose, they were removed from their original growth substrate by gentle sonication for a few seconds in isopropanol, and the resulting solution was poured on an Au-evaporated/Si substrate kept on a hot plate for solvent evaporation. Collected photons were dispersed through a 30 cm focal length monochromator and detected by a Hamamatsu RS374 detector. CL spectra were recorded by scanning over the entire nanowire area. During spectral acquisition, the photon integration time for each acquisition wavelength was carefully chosen as an integer multiple of the beam scan time, which avoids possible artifacts in the spectra due to spatial inhomogeneity in the sample emission.

High-angle annular dark-field STEM (HAADF-STEM) tomography was employed to obtain the 3D reconstruction of a particular GaAs-AlGaAs core-multishell nanowire (from Sample B). To this end, an HAADF-STEM image tilt series within an angular range of ±68° and in steps of 2° was recorded on an FEI Titan 80-300 microscope operated at an electron acceleration voltage of 300 kV. The 1000×1000 HAADF-STEM images with a pixel size of 0.97 nm were then aligned, i.e. corrected for nanometer displacements between subsequent projections within the tilt series, and finally reconstructed using weighted simultaneous iterative reconstruction technique [17] yielding the nanowire tomogram. So-called missing wedge artifacts in the tomogram, caused by the incomplete tilt range available (±68° instead of ±90°), lead to a reduced resolution in the corresponding directions. This problem was reduced by picking a nanowire resting with one of its ⟨112⟩ edges (corners of the hexagonal cross-section) on the carbon support of the TEM grid. The reason is that, in this orientation, the faces (edges in the hexagonal cross-section) are parallel to the ±60° projection directions and hence better resolved. Further details about the HAADF-STEM tomography experiment and reconstruction can be found in Ref. [18].

## 3 Results and discussion

Figure 1A–C illustrates the CL spectra recorded at 7 K for three nanowires among those selected (from Sample A). The spectra in Figure 1A and B refer to nanowires having almost the same lengths (3.45 and 3.67 μm, respectively) but slightly different diameters. Experimental spectra

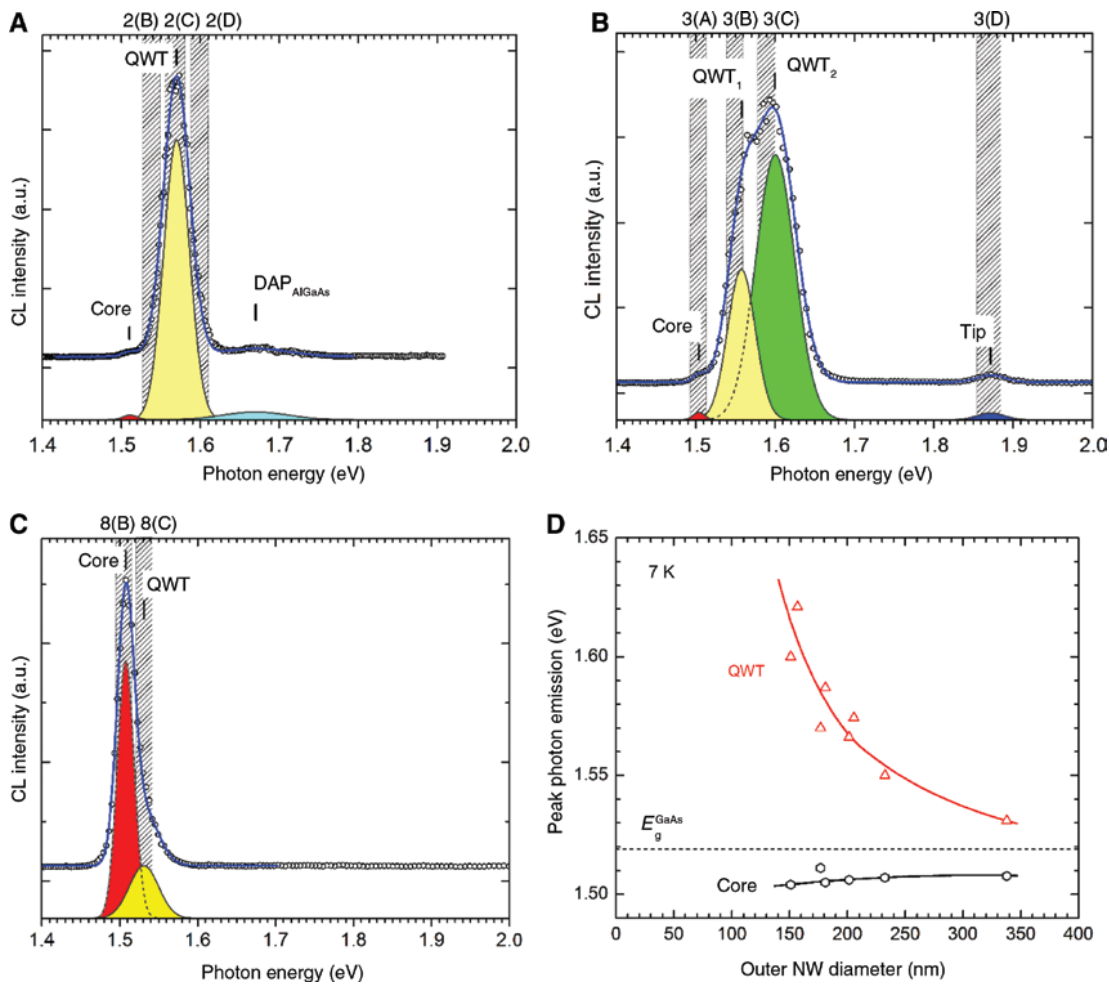

**Figure 1:** (A–C) Experimental CL spectra (O) recorded at 7 K for three different GaAs-AlGaAs core-multishell nanowires from Sample A (see Figures 2, 3, and 8 for corresponding FE-SEM and CL images).
The solid blue line in each diagram is the line-shape best-fitting of experimental data by two or more Gaussian peaks, accounting for various contributions, namely the core, QWT, and nanowire tip (if any), as indicated in the diagrams: best-fitting peaks (gray solid lines) are shown offset below the experimental spectrum for clarity sake. The gray-hatched vertical regions in each diagram indicate the photon spectral intervals (~20 meV wide) employed for collecting the energy-selected CL images in Figures 2, 3, and 8, as indicated above the diagrams.
(D) Best-fitting peak energy values for the core and QWT emissions as a function of nanowire diameter (solid curves are only guides for the eye).

were line-shape best-fitted with two or more Gaussian peaks, to account for the various nanowire contributions, namely the GaAs inner core, the QWT, and those related to AlGaAs point defects or other structural features (if any). The spectrum in Figure 1A shows a dominant emission peaked at 1.570 eV, i.e. well above the band-gap energy (Eg(GaAs)=1.519 eV at 7 K) of GaAs, which can be ascribed to the GaAs QWT within the nanowire [5]. Upon deconvolution of the fitted peak full-width at half-maximum (FWHM) by the monochromator instrumental width, the intrinsic broadening of the QWT emission turns out to be about 37.5 meV. A much weaker peak contribution also appears at 1.511 eV, i.e. within the spectral range where exciton recombination from the GaAs core is usually observed [19, 20], along with a weak and broader band at about 1.67 eV; the latter has been observed in the low-temperature luminescence of GaAs-AlGaAs core-shell nanowires and ascribed to the donor-acceptor pair recombination in the AlGaAs alloy [15]. Panchromatic imaging of this nanowire (Figure 2A) shows that the recorded CL emission mostly originates from a 2.3-μm-long section of the nanowire close to its tip, the remaining part (between 2.3 and 3.3 μm from the tip; see Figure 2F) appearing dark in the CL image, likely as a result of structural defects within the nanowire. Figure 2B–D shows the energy-selected CL images recorded for three distinct spectral intervals (indicated by the gray-hatched vertical regions in Figure 1A) centered around the QWT peak energy and its low- and high-energy shoulders: noteworthy, as the selected energy intervals moves from lower to higher energies, the emission intensity in the CL images tends to slightly shift along the nanowire axis, from regions close to the tip toward the nanowire center. As the total nanowire diameter slightly shrinks from tip (180 nm) to center (175 nm; see Figure 2F), one would also expect for the actual GaAs shell thickness, and this affects its emission energy contributing to the observed broadening of the QWT peak. These findings are further evidenced in the nanowire of Figures 1B and 3: besides the weak exciton contribution (1.504 eV) from the core, the CL spectrum shows a dominant and broad band between 1.52 and 1.67 eV, which can be accounted for by two Gaussian contributions peaked at 1.558 eV ($QWT_1$) and 1.600 eV ($QWT_2$). The energy-selected CL images in Figure 3B and C show that the $QWT_1$ emission originates from a 0.6-µm-long region of the nanowire close to its tapered tip and (to a much lower extent) from a 0.3-µm-long section toward the base, whereas the $QWT_2$ emission comes predominantly from the center of the nanowire trunk. Figure 3E further demonstrates that the $QWT_2$ emission is associated with a nanowire region of fairly constant diameter (151 nm), whereas the $QWT_1$ peak arises from nanowire sections of slightly larger diameters, supporting again a strict correlation between the QWT peak energy and its local (i.e. along the nanowire) thickness. In comparison, the core emission arises from the entire nanowire length (Figure 3A), its intensity increasing however, from the nanowire base to its tip. Finally, the weak emission at 1.872 eV originates from a small region within the nanowire tapered tip section (Figure 3D). We ascribe this latter emission to the growth of graded-alloy AlGaAs core sections during the deposition of GaAs shells as a result of the VLS growth induced by the Al reservoir left in the Au catalyst nanoparticle during the previous AlGaAs shell growth [16, 21]. Figure 1C reports instead the CL spectrum of a much larger nanowire (388 nm). In this case, the spectral intensities were best fitted by two Gaussian contributions: a dominant peak at 1.508 eV, due to the core, plus a weaker and larger emission at about 1.531 eV ascribed to the QWT. Interestingly, the core emission dominates over the QWT due to the large core diameter (106 nm) of this nanowire. In this respect, the core diameters of the nanowires in Figures 2 and 3 were about 63 and 54 nm, respectively. Energy-selected CL images recorded for this nanowire are reported in Figure 8 and will be discussed further below.

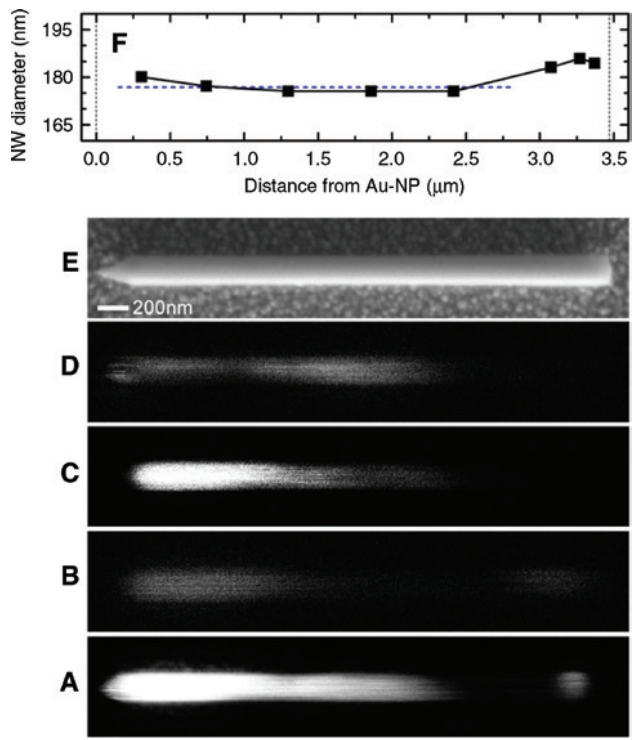

**Figure 2:** (A) Panchromatic CL image of the nanowire whose spectrum is reported in Figure 1A, (B–D) energy-selected CL images corresponding to the spectral intervals evidenced by the gray-hatched vertical regions in Figure 1A, (E) FE-SEM micrograph of the nanowire (the 200 nm marker in the micrograph holds also for CL images), and (F) values of nanowire diameters measured by FE-SEM at different positions along its length.
The dashed horizontal line spans a relatively uniform diameter region corresponding to an average value of 177 nm.

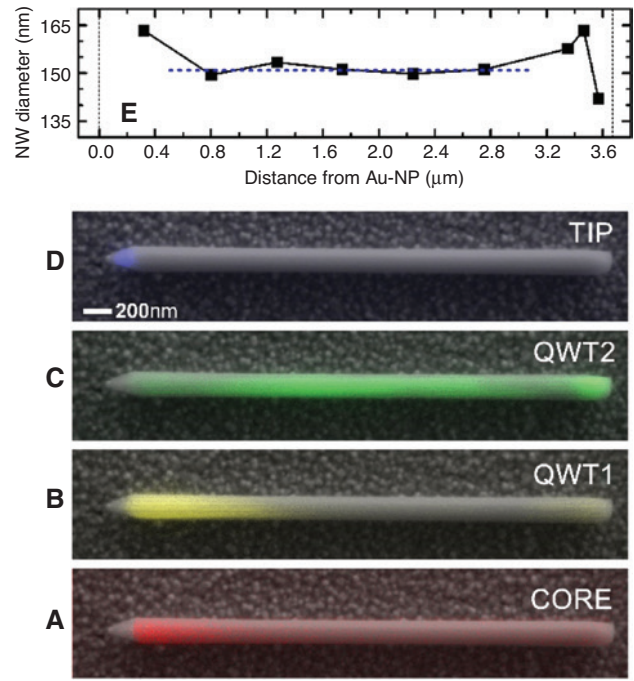

**Figure 3:** (A–D) Energy-selected CL images corresponding to the spectral intervals evidenced by the gray-hatched vertical regions in Figure 1B; the nanowire FE-SEM image has been superposed to the CL images to evidence the spatial origin of each CL contribution. The 200 nm marker in (D) holds for all images. (E) Values of nanowire diameters measured by FE-SEM at different positions along its length. The dashed horizontal line spans a relatively uniform diameter region corresponding to an average value of 151 nm.

Figure 1D reports the as-determined core and the QWT emission peak energies for various nanowires as a function of their diameters; in the figure, care was taken to select only those QWT emissions that could be clearly associated – through the analysis of the corresponding CL images – with nanowires, or nanowire sections, having relatively uniform (within a few percent; see Figures 2F and 3E) diameters. Noteworthy, the core energy only slightly changes with the nanowire diameter, likely as a result of small changes of its built-in strain [20], as the core and overgrown shells change their diameter/thickness. However, the QWT energy tends to greatly increase with decreasing the overall nanowire diameter, indicating that the average thickness of each GaAs QWT also decreases: clearly, one would like to correlate the QWT emission energy with its actual thickness, and this requires to somehow extract the GaAs shell thickness within each of the selected nanowires in Figure 1D.

In the following, we demonstrate that this is possible based on the knowledge of the peculiar shell growth mode of present nanowires, which allows to quantitatively predict the thickness of the GaAs QWT within each nanowire, based on their FE-SEM measured core and overall diameters: indeed, as the nanowires in Figure 1D belong to the same sample, their different diameters imply different radial growth rates for QWTs belonging to different nanowires. This effect was previously reported for AlGaAs shells grown around GaAs cores in dense nanowire arrays and explained as a result of the group III vapor mass-transport limited growth regime holding during high-temperature MOVPE [22]. Given an array of free-standing nanowires on the substrate, the group III precursor flux arriving at the growth front during the shell growth time will divide between the free substrate surface and the nanowire array cumulative lateral surface: the application of the law of mass conservation, under the assumption of (i) homogeneous vapor concentration of group III precursors along the nanowire height and (ii) perfect conformality of material overgrowth between the substrate and nanowires, leads to write a differential equation expressing the shell radial growth rate (namely $dh_s/dt$) as [22]:

$$\frac{dh_s}{dt} \equiv \frac{1}{2}\frac{dD_{NW}}{dt} = \frac{R^0_{2D}}{1+2\sqrt{3}D_{NW}L_{NW}\delta_{NW}} \quad (1)$$

where $D_{NW}$ is the total nanowire diameter measured normal to the nanowire [110] facets, $\delta_{NW}$ is the local (on the substrate) nanowire areal density, $L_{NW}$ is the average nanowire length within the array, and $R^0_{2D}$ is the planar growth rate of the material. Therefore, as the $L_{NW}\delta_{NW}$ product varies across the sample, the shell growth rate around the nanowires will change. Equation (1) can be readily integrated over the growth time to calculate the actual shell thickness and predicts a nonlinear dependence on the actual value of $L_{NW}\delta_{NW}$, initial GaAs nanowire diameters (proportional to the Au catalyst nanoparticle size), and deposition time. Further details are reported in Ref. [22].

The shell growth model above can be generalized to the case of multiple shell nanowires [23] and in particular employed to predict the thickness of the GaAs QWT within the nanowire. To this purpose, the generalized model is here briefly validated against experimental data (core and nanowire diameters and AlGaAs and GaAs shell thickness) provided by HAADF-STEM tomographic reconstruction of a particular core-multishell nanowire (from Sample B) [18]. Figure 4 reports the 3D analysis of the obtained HAADF-STEM tomogram in terms of the nanowire internal structure provided by the different Z-contrast of its constituent GaAs and $Al_{0.33}Ga_{0.67}As$ shells (Figure 4A). It reveals local variations in the shell thicknesses in axial, radial, and azimuthal directions within a range of a few nanometers as analyzed statistically further below. Figure 4B–D demonstrates local measurements of core and shell thicknesses by investigating 1D profiles across the layers taken at line-scans through a [110] slice. We repeated this procedure for the other two [110] and three [112] nanowire directions.

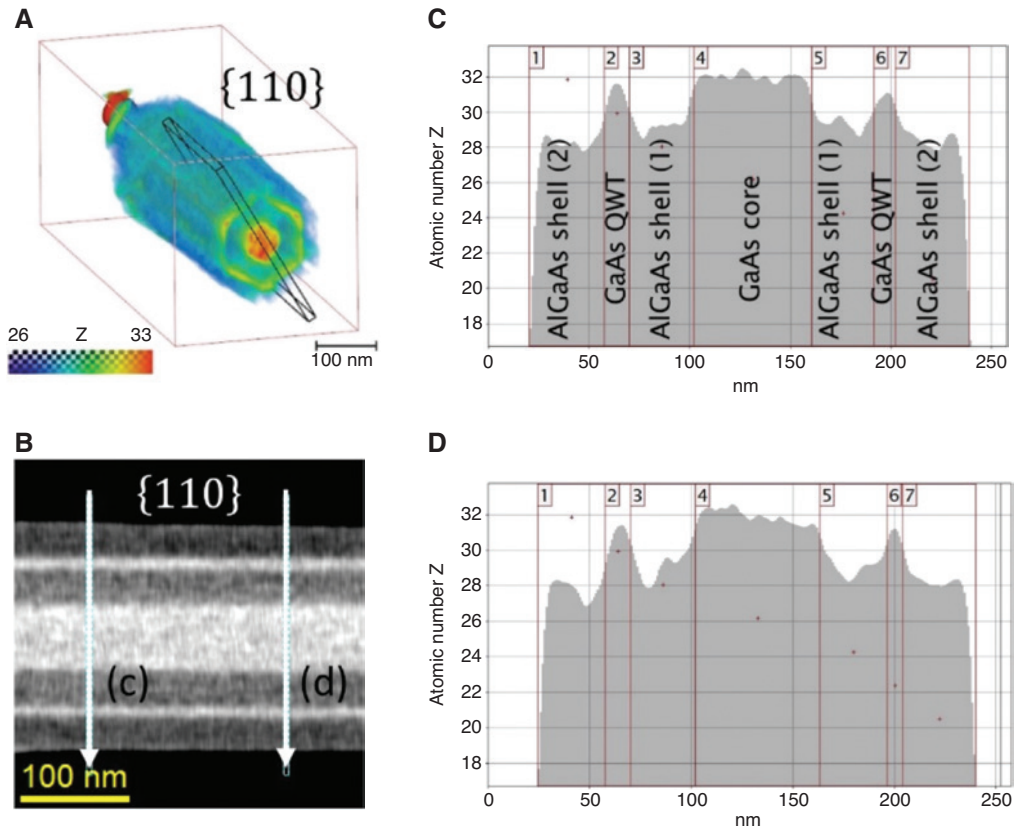

**Figure 4:** HAADF-STEM (Z-contrast) tomography on a GaAs-AlGaAs core-multishell nanowire.
(A) 3D volume rendering of the nanowire tomogram with atomic number Z color-coded, i.e. the GaAs regions (core and QWT) are visualized as red-yellow and the AlGaAs shell as green-blue. (B) Extracted 2D slice along one [110] direction at the trunk region of the nanowire at position indicated by the black frame in (A). It shows the GaAs core and QWT in white and the AlGaAs shells in gray. (C and D) Profiles in cross-sectional direction along the line scans in (B) marked by the white arrows and their assignment to the different layers in (C).

The as-determined average core/nanowire diameters and the thickness values for each AlGaAs and GaAs shell in Figure 4 are reported in Table 1, but the GaAs cap shell which remained difficult to estimate due to its relatively small value (close to HAADF-STEM tomography spatial resolution) and rather inhomogeneous morphology [18]. In Table 1, data are compared to values predicted by the multishell growth model using the shell growth times and the measured nanowire core diameter as input values.[1] Best agreement between experimental and calculated data was obtained for $L_{NW}\delta_{NW} = 1.2 \times 10^{-3}$ nm$^{-1}$; noteworthy, if the 4 μm length of the nanowire in Figure 4 would be representative of the average length within the array, the nanowire density can then be estimated as $\delta_{NW} \approx 3 \times 10^{7}$ cm$^{-2}$, i.e. well within the experimental range for present samples. In Table 1, the best-fitted nanowire diameter and shell thicknesses appear very close (within experimental errors in most cases) to experimental values; noteworthy, the calculated QWT thickness differs from the one measured by about 0.4 nm (i.e. ~1 monolayer of GaAs in the [110] direction). More details on the multishell growth model will be reported elsewhere [23].

---

[1] Taking $R^{0}_{2D}(GaAs) = 0.122$ nm/s and $R^{0}_{2D}(AlGaAs) = 0.358$ nm/s as planar growth rates for GaAs and AlGaAs, respectively, under employed MOVPE conditions.

**Table 1:** Comparison between GaAs and AlGaAs shell thickness values, as measured from analysis of HAADF-STEM tomography of the GaAs-AlGaAs core-multishell nanowire in Figure 4, and simulated ones based on the multishell growth model [22, 23].

| Sample B | NW diameter (nm) | | Shell thickness[a] (nm) | | | | $L_{NW}\delta_{NW}$ ($10^{-3}$ nm$^{-1}$) |
|---|---|---|---|---|---|---|---|
| | Core | Total | First AlGaAs shell | GaAs QWT | Second AlGaAs shell | GaAs cap | |
| TOMO-STEM | 60.7±1.5 | 215.7±3.8 | 33.0±1.3 | 9.7±2.2 | 35.0±1.4 | N/A | – |
| Multishell growth model | 60.7[b] | 219.6 | 31.1 | 10.1 | 34.4 | 3.9 | 1.2 |

[a]Along the [110] directions.
[b]Input value.
$L_{NW} \cdot \delta_{NW}$ is the model free-parameter value (see Discussion) which best-fits experimental data.

Based on the growth model above, we then estimated the AlGaAs and GaAs shell thickness for each of the nanowires reported in Figure 1D using the shell growth times and the FE-SEM measured core and nanowire diameters as input values; in particular, the core diameters were taken equal to the size of the Au-catalyst particle at the nanowire tip. Calculated data are reported in Figure 5A as a function of the best-fitting $L_{NW}\delta_{NW}$ value for each nanowire. As their average length is 3.5 μm, this indicates that the nanowires selected for CL characterization came from areas of the sample with nanowire densities varying between $1.2 \times 10^6$ and $2.9 \times 10^8$ cm$^{-2}$.

Finally, Figure 5B reports the energy of the QWT peak emission as a function of as-calculated GaAs shell thickness (between 3 and 7 nm) for the analyzed nanowires. As expected, the QWT peak energy blue-shifts with decreasing the GaAs thickness; however, our experimental values remain 40–120 meV below the theoretical calculations reported in Ref. [6], the latter representing exciton emission energies due to the recombination of the first ($e_1$) and second ($e_2$) confined electron states to the ground ($hh_1$) heavy-hole state within a uniform GaAs QWT symmetrically wrapped around the core between two $Al_{0.35}Ga_{0.65}As$ shells.

QWT asymmetries (between facets) and thickness fluctuations could lead to pronounced carrier localization at regions of the QWT where its thickness is largest [12], causing substantial red-shifts of the exciton emission. To evidence the occurrence of asymmetries and/or thickness fluctuations in present GaAs shells, we performed a statistical analysis on the STEM tomogram (Z-contrast) of a section of the nanowire trunk in Figure 4. The axial length (in the z-direction) of the analyzed region is about 320 nm consisting of 66 slices of 4.9 nm width. At each slice, the GaAs QWTs were segmented (binarized) by applying a threshold of $Z = 31.2$ (corresponding to 10% Al content).

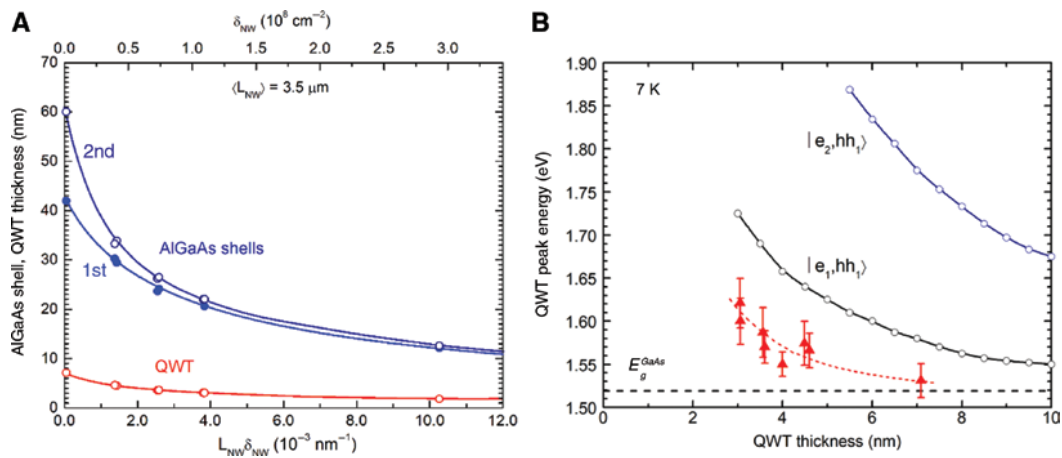

**Figure 5:** (A) Average GaAs QWT and AlGaAs shell thickness within each nanowire as a function of their $L_{NW}\delta_{NW}$ value, calculated based on the multishell nanowire growth model (Sample A).
Solid curves are only guides for the eye. (B) Peak energy of QWT emissions as a function of QWT thickness: experimental values (▲) were determined from Gaussian line-shape best-fitting of CL spectra (see Figure 1), whereas QWT thickness are from data in (A). Error bars represent the peak intrinsic broadening. Open symbols (○) represent calculated $|e_1,hh_1\rangle$ and $|e_2,hh_1\rangle$ exciton transition energies for constant thickness QWTs grown in between two $Al_{0.35}Ga_{0.65}As$ barrier shells (after Ref. [6]).

To account for noise and limited resolution of experimental data and distinguish between GaAs and AlGaAs shells along the six [110] and [112] directions, the next nearest-neighbor pixels of those belonging to the GaAs core and QWT, were included to GaAs regions if they were higher than $Z = 30.2$ (corresponding to 20% Al content); otherwise, they were added to AlGaAs regions. To determine the azimuthal thickness variations of the QWT, their binary data were integrated (projected) perpendicular to the QWT, i.e. along the six [110] and [112] directions. The thickness histograms calculated from the projection data revealed the thickness variations of the QWT (Figure 6). The 12 histograms in the figure show relatively broad peaks (FWHM ~4 nm) over the 5–15 nm thickness interval. The 0 nm counts in the histograms can be mainly attributed to missing wedge artifacts but might also suggest discontinuities in the QWT. The tomographic data exhibit fluctuations not only in the thickness but also in the signal height of the QWT. The latter can be interpreted either as local concentration changes, i.e. Al intermixing, or as an average effect if the QWT is thinner than the measured region in the 3D volume (or even smaller than the reconstructed voxel size of the tomogram). This is exemplified in Figure 7, where the Z-contrast in the STEM tomogram along two slices parallel to the [110] or [112] planes is evaluated.

Noteworthy, the QWT thickness values along the [110] direction are consistent (within experimental uncertainties) with that reported in Table 1, whereas its value is larger (average thickness ~14 nm) along three of the six [112] directions and smaller (~8 nm) in the opposite directions, giving the QWT a peculiar trigonal symmetry. Figure 6 further shows that, within each trigonal sector, the QWT takes a crescent-shaped morphology.

Clearly, the confined carrier wave-functions will localize where the QWT width is largest [7, 12], i.e. at the three thicker [112] corners of the GaAs shell (red-colored areas in Figure 6), leading to a decrease of their confined-state energies. Such localized electron states would run along the length of the nanowire, showing a 1D wire-like character [7]; however, a combination of azimuthal and axial fluctuations of the QWT thickness may give its quantum-confined states a 0D dot-like character: the latter states were shown to occur in narrow (<5 nm) GaAs QWTs within GaAs-AlGaAs core-multishell nanowires as a result of QWT thickness or alloy fluctuations along the nanowire [11].

The occurrence of thickness fluctuations in present QWTs can be directly evidenced by CL imaging of the nanowire emissions. Figure 8B and C reports energy-selected images corresponding to the nanowire core and QWT emissions in Figure 1C, for which our multishell growth model predicts a 7.1-nm-thick GaAs QWT grown between a 42.0 nm inner AlGaAs shell and a 60.0 nm outer AlGaAs shell (see Figure 5A). Noteworthy, the core and QWT emission channels appear radially well-resolved, the core emission intensity predominantly arising from the central nanowire region (approximately corresponding to the core volume), whereas the QWT emission comes from two thin regions radially offset from the nanowire center toward its surface. This effect is more easily visualized in the color-coded superposition of the two images above (Figure 8D). As the nanowire appears resting with one of its ⟨112⟩ edges on the Au/Si substrate (Figure 8A), during CL measurements, the microscope primary electron beam remained parallel to two of the nanowire [110] facets and thus preferentially excited those volumes of the QWT shell parallel to these facets; this fact, along with the nanowire relatively thick AlGaAs shells, allowed to spatially separate the two emissions. More interestingly, whereas the core emission shows a relatively uniform CL intensity throughout the entire nanowire length (Figure 8B), the QWT emission appears localized at specific positions along the nanowire (Figure 8C); furthermore, these positions do not replicate symmetrically on either sides of the core (i.e. above and below the core region in the figure), demonstrating that the localization effect is also azimuth dependent. These findings allow to state that the 40–120 meV red-shift of the QWT emission reported in Figure 5B can indeed be ascribed to spatial localization of confined carrier at thickness fluctuations of the GaAs QWT. It is expected that these localized states would show a more 1D- or 0D-like character depending on the width of the QWT. Further investigations are thus required to clarify this aspect for present nanowire structures.

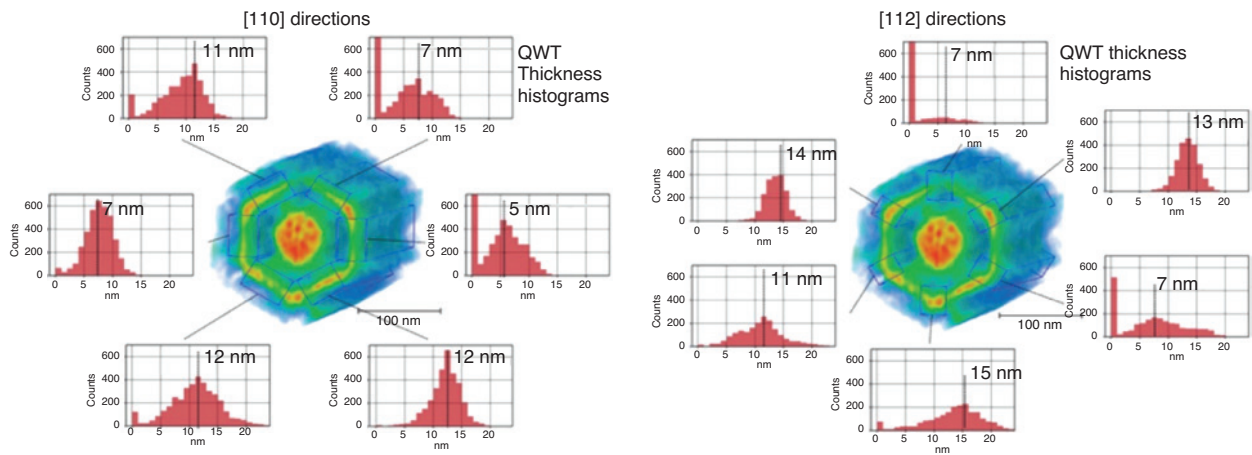

**Figure 6:** Analysis of the QWT thickness distribution of a trunk section of the GaAs-AlGaAs core-multishell nanowire in Figure 4 along the [110] and [112] directions.
The thickness histograms were computed within the regions (320 nm long along the nanowire axial direction) indicated by the blue boxes in the volume rendering of the tomogram.

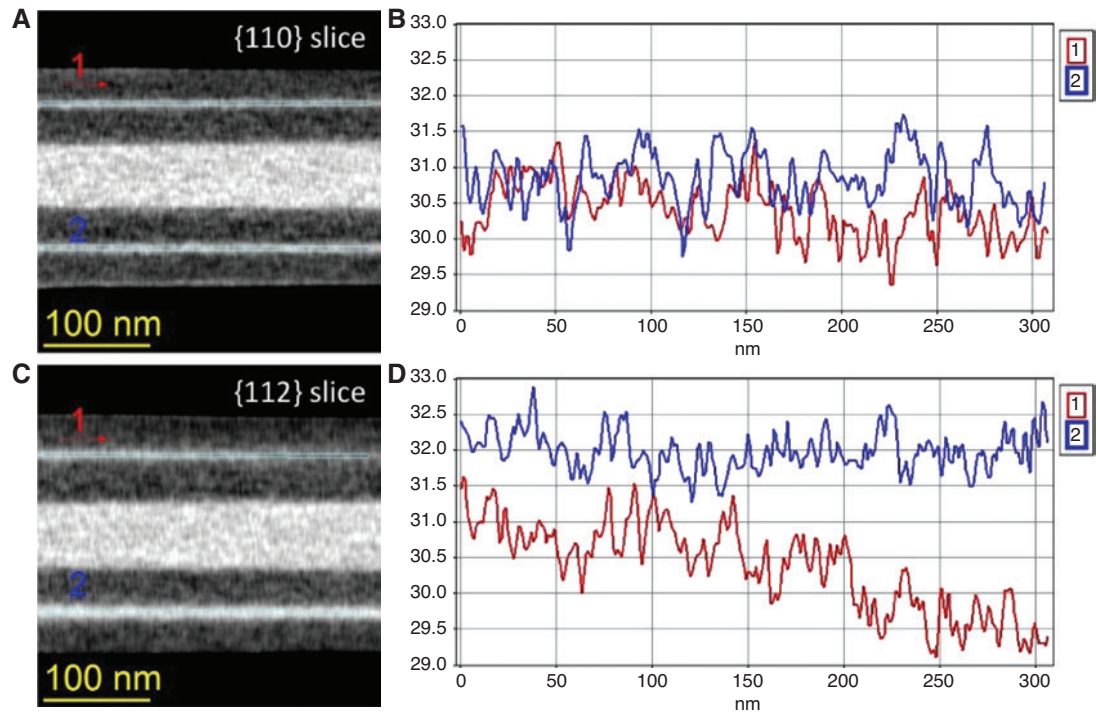

**Figure 7:** (A and C) Axial fluctuations in GaAs content (i.e. Al intermixing) within the QWT of the GaAs-AlGaAs core-multishell nanowire reconstructed by HAADF-STEM tomography (Figure 4) along slices parallel to the [110] or [112] planes.
(B and D) Line profiles (1,2) taken at line scans inside the QWT as indicated by the 1 and 2 label in (A and C). They show deviations from $Z=32$ (average atomic number of GaAs), indicating that the concentration varies locally: for example, in (D), the line profile 1 goes down until $Z=29.5$, which corresponds to about 30% of Al.

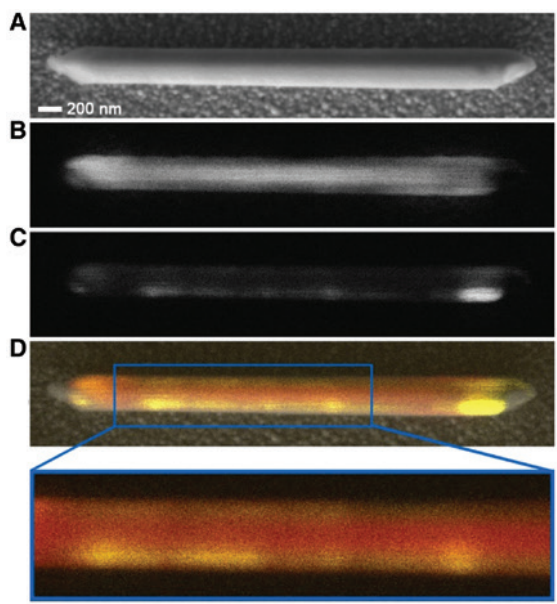

**Figure 8:** (A) FE-SEM micrograph of the nanowire whose spectrum is reported in Figure 1C (the 200 nm marker holds also for CL images); (B and C) energy-selected CL images corresponding to the spectral intervals evidenced by the gray-hatched vertical regions in Figure 1C; and (D) color-coded (red: core; yellow: QWT) superposition of the two CL images in (B) and (C), allowing a better visualization of the spatial localization of the QWT emission.

# 4 Conclusion

We reported on the nanoscale luminescence emission of GaAs-AlGaAs QWT nanowires by 7 K CL spectroscopy and imaging performed on single nanowire structures. For each nanostructure, contributions to its luminescence from the core, the QWT, and AlGaAs point defects or minor structural features (if any) were identified. In particular, CL imaging of QWT emissions allowed to ascribe each contribution to a specific region along the nanowire, associated with a well-defined diameter value, the latter determined by FE-SEM inspection. It appeared that the QWT peak energy strongly blue-shifts with decreasing the diameter of the nanowire region originating the emission, an effect ascribed to the different inner structure (i.e. QWT thickness) of selected nanowires.

To extract the actual shell thickness values corresponding to a specific nanowire region, we applied a multishell growth model, which dictates the actual growth rates of coaxial GaAs and AlGaAs shells around a dense array of free-standing core nanowires depending on the local (on the growth substrate) nanowire length ($L_{NW}$) and density ($\delta_{NW}$). The model was successfully validated (through best-fitting) against experimental shell thickness data obtained from the analysis of the [110] slices of the 3D reconstructed HAADF-STEM tomogram of a particular GaAs-AlGaAs QWT nanowire; in this case, the calculated QWT average thickness (10.1 nm) well reproduced (within experimental uncertainties) the measured value (9.7 nm) with a best-fitting parameter $L_{NW}\delta_{NW} = 1.2 \times 10^{-3}$ nm$^{-1}$. Statistical analysis of the QWT along the [110] and [112] slices of the nanowire 3D tomogram further evidenced a prominent azimuthal asymmetry and fluctuation of the GaAs shell thickness, the QWT being thicker than 9.7 nm at three of its [112] hexagonal corners and thinner at the other three (opposite) corners. In addition, the QWT thickness randomly fluctuates along the nanowire axis.

The peak energies of QWT emissions as a function of as-determined GaAs shell thickness turned out to be 40–120 meV below the theoretical calculations reported in the literature for exciton recombination involving carriers confined within uniform GaAs QWTs symmetrically wrapped around the core. As-estimated red-shifts were ascribed to carrier localization at the thick [112] corners of the GaAs QWT, their confined states being expected to have a 1D (wire-like) or 0D (dot-like) character depending on the actual combination of azimuthal and axial changes/fluctuations of the QWT width within the nanowire. Finally, CL mapping of the QWT luminescence along the nanowire axis allowed to directly image nanoscale localizations of the emission, strongly supporting the case of carrier localization in our QWT nanowires.

**Acknowledgments:** The authors would like to acknowledge E. Stevanato for her contribution to the MOVPE growth of the nanowire samples analyzed in this work and R. Rosato for her collaboration with CL measurements.